\documentstyle[12pt]{article} 
\catcode`\@=11

\def\psfortextures{
\def\PSspeci@l##1##2{%
\special{illustration ##1\space scaled ##2}}}

\def\psfordvips{
\def\PSspeci@l##1##2{%
\d@my=0.1bp \d@mx=\drawingwd \divide\d@mx by\d@my%
\includegraphics{##1\space}}}

\def\psforoztex{
\def\PSspeci@l##1##2{%
\special{##1 \space
      ##2 1000 div dup scale
      \putsp@ce{\number-\psllx} \putsp@ce{\number-\pslly} translate}}}
\def\putsp@ce#1{#1 }

\def\psonlyboxes{
\def\PSspeci@l##1##2{%
\at{0cm}{0cm}{\boxit{\vbox to\drawinght
  {\vss
  \hbox to\drawingwd{\at{0cm}{0cm}{\hbox{(##1)}}\hss}
  }}}}}

\newdimen\drawinght\newdimen\drawingwd
\newdimen\psxoffset\newdimen\psyoffset
\newbox\drawingBox

\newread\epsffilein    
\newif\ifepsffileok    
\newif\ifepsfbbfound   
\newif\ifepsfverbose   
\newdimen\epsfxsize    
\newdimen\epsfysize    
\newdimen\epsftsize    
\newdimen\epsfrsize    
\newdimen\epsftmp      
\newdimen\pspoints     
\def\ReadPSize#1{
\edef\PSfilename{#1}
\global\def\epsfllx{72}
\global\def\epsflly{72}
\global\def\epsfurx{540}
\global\def\epsfury{720}
\openin\epsffilein=#1
\ifeof\epsffilein\errmessage{I couldn't open #1, will ignore it}\else
   {\epsffileoktrue \chardef\other=12
    \def\do##1{\catcode`##1=\other}\dospecials \catcode`\ =10
    \loop
       \read\epsffilein to \epsffileline
       \ifeof\epsffilein\epsffileokfalse\else
          \expandafter\epsfaux\epsffileline:. \\%
       \fi
   \ifepsffileok\repeat
   \ifepsfbbfound\else
    \ifepsfverbose\message{No bounding box comment in #1; 
using defaults}\fi\fi
   }\closein\epsffilein\fi
\def\psllx{\epsfllx}\def\pslly{\epsflly}%
\def\psurx{\epsfurx}\def\psury{\epsfury}%
\drawinght=\epsfury bp%
\advance\drawinght by-\epsflly bp%
\drawingwd=\epsfurx bp%
\advance\drawingwd by-\epsfllx bp%
}

{\catcode`\%=12 \global\let\epsfpercent=
\long\def\epsfaux#1#2:#3\\{\ifx#1\epsfpercent
   \def\testit{#2}\ifx\testit\epsfbblit
      \epsfgrab #3 @ @ @ \\%
      \epsffileokfalse
      \global\epsfbbfoundtrue
   \fi\else\ifx#1\par\else\epsffileokfalse\fi\fi}%

\def\epsfgrab #1 #2 #3 #4 #5\\{%
   \global\def\epsfllx{#1}\ifx\epsfllx\empty
      \epsfgrab #2 #3 #4 #5 @\\\else
   \global\def\epsflly{#2}%
   \global\def\epsfurx{#3}\global\def\epsfury{#4}\fi}%

\newcount\xscale \newcount\yscale \newdimen\pscm\pscm=1cm
\newdimen\d@mx \newdimen\d@my
\let\ps@nnotation=\relax
\def\psboxto(#1;#2)#3{\vbox{
   \catcode`\:=12
   \ReadPSize{#3}
   \divide\drawingwd by 1000
   \divide\drawinght by 1000
   \d@mx=#1
   \ifdim\d@mx=0pt\xscale=1000
         \else \xscale=\d@mx \divide \xscale by \drawingwd\fi
   \d@my=#2
   \ifdim\d@my=0pt\yscale=1000
         \else \yscale=\d@my \divide \yscale by \drawinght\fi
   \ifnum\yscale=1000
         \else\ifnum\xscale=1000\xscale=\yscale
                    \else\ifnum\yscale<\xscale\xscale=\yscale\fi
              \fi
   \fi
   \divide \psxoffset by 1000\multiply\psxoffset by \xscale
   \divide \psyoffset by 1000\multiply\psyoffset by \xscale
   \global\divide\pscm by 1000
   \global\multiply\pscm by\xscale
   \multiply\drawingwd by\xscale \multiply\drawinght by\xscale
   \ifdim\d@mx=0pt\d@mx=\drawingwd\fi
   \ifdim\d@my=0pt\d@my=\drawinght\fi
\message{[#3\space [BoundingBox\string:
\space\epsfllx\space\epsflly\space\epsfurx\space\epsfury]}%
\message{[scaled\space\the\xscale\string:
\space\the\drawingwd\space x \the\drawinght]]}%
 \hbox to\d@mx{\hss\vbox to\d@my{\vss
   \global\setbox\drawingBox=\hbox to 0pt{\kern\psxoffset\vbox to 0pt{
      \kern-\psyoffset
      \PSspeci@l{\PSfilename}{\the\xscale}
      \vss}\hss\ps@nnotation}
   \global\ht\drawingBox=\the\drawinght
   \global\wd\drawingBox=\the\drawingwd
   \baselineskip=0pt
   \copy\drawingBox
 \vss}\hss}
  \global\psxoffset=0pt
  \global\psyoffset=0pt
  \global\pscm=1cm
  \global\drawingwd=\drawingwd
  \global\drawinght=\drawinght
}}

\def\psboxscaled#1#2{\vbox{
  \catcode`\:=12
  \ReadPSize{#2}
  \xscale=#1
  \divide\drawingwd by 1000\multiply\drawingwd by\xscale
  \divide\drawinght by 1000\multiply\drawinght by\xscale
  \divide \psxoffset by 1000\multiply\psxoffset by \xscale
  \divide \psyoffset by 1000\multiply\psyoffset by \xscale
  \global\divide\pscm by 1000
  \global\multiply\pscm by\xscale
\message{[#2\space [BoundingBox\string:
\space\epsfllx\space\epsflly\space\epsfurx\space\epsfury]}%
\message{[scaled\space\the\xscale\string:
\space\the\drawingwd\space x \the\drawinght]]}%
  \global\setbox\drawingBox=\hbox to 0pt{\kern\psxoffset\vbox to 0pt{
     \kern-\psyoffset
     \PSspeci@l{\PSfilename}{\the\xscale}
     \vss}\hss\ps@nnotation}
  \global\ht\drawingBox=\the\drawinght
  \global\wd\drawingBox=\the\drawingwd
  \baselineskip=0pt
  \copy\drawingBox
  \global\psxoffset=0pt
  \global\psyoffset=0pt
  \global\pscm=1cm
  \global\drawingwd=\drawingwd
  \global\drawinght=\drawinght
}}

\def\psannotate#1#2{\def\ps@nnotation{#2\global\let\ps@nnotation=\relax}#1}
\def\pscaption#1#2{\vbox{
   \setbox\drawingBox=#1
   \copy\drawingBox
   \vskip\baselineskip
   \vbox{\hsize=\wd\drawingBox\setbox0=\hbox{#2}
     \ifdim\wd0>\hsize
       \noindent\unhbox0\tolerance=5000
    \else\centerline{\box0}
    \fi
}}}

\def\at#1#2#3{\setbox0=\hbox{#3}\ht0=0pt\dp0=0pt
  \rlap{\kern#1\vbox to0pt{\kern-#2\box0\vss}}}

\newdimen\gridht \newdimen\gridwd
\def\gridfill(#1;#2){
  \setbox0=\hbox to 1\pscm
  {\vrule height1\pscm width.4pt\leaders\hrule\hfill}
  \gridht=#1
  \divide\gridht by \ht0
  \multiply\gridht by \ht0
  \advance \gridht by \ht0
  \gridwd=#2
  \divide\gridwd by \wd0
  \multiply\gridwd by \wd0
  \advance \gridwd by \wd0
  \vbox to \gridht{\leaders\hbox to\gridwd{\leaders\box0\hfill}\vfill}}

\def\frameit#1#2#3{\hbox{\vrule width#1\vbox{
  \hrule height#1\vskip#2\hbox{\hskip#2\vbox{#3}\hskip#2}%
        \vskip#2\hrule height#1}\vrule width#1}}
\def\boxit#1{\frameit{0.4pt}{0pt}{#1}}

\catcode`\@=12 

\psfordvips

\jot = 1.5ex
\def\baselinestretch{1.65}
\parskip 5pt plus 1pt

\catcode`\@=11

\@addtoreset{equation}{section}

\def\@normalsize{\@setsize\normalsize{15pt}\xiipt\@xiipt
\abovedisplayskip 14pt plus3pt minus3pt%
\belowdisplayskip \abovedisplayskip
\abovedisplayshortskip  \z@ plus3pt%
\belowdisplayshortskip  7pt plus3.5pt minus0pt}
\def\small{\@setsize\small{13.6pt}\xipt\@xipt
\abovedisplayskip 13pt plus3pt minus3pt%
\belowdisplayskip \abovedisplayskip
\abovedisplayshortskip  \z@ plus3pt%
\belowdisplayshortskip  7pt plus3.5pt minus0pt
\def\@listi{\parsep 4.5pt plus 2pt minus 1pt
            \itemsep \parsep
            \topsep 9pt plus 3pt minus 3pt}}

\def\underline#1{\relax\ifmmode\@@underline#1\else
        $\@@underline{\hbox{#1}}$\relax\fi}
\@twosidetrue
\relax

\catcode`@=12

\evensidemargin 0.0in
\oddsidemargin 0.0in
\topmargin -0.2in
\textwidth 6.4in
\textheight 8.9in

\catcode`\@=11

\def\section{\@startsection{section}{1}{\z@}{3.5ex plus 1ex minus
   .2ex}{2.3ex plus .2ex}{\large\bf}}

\def\ps@headings{\def\@oddfoot{}\def\@evenfoot{}
\def\@oddhead{\hbox{}\hfill
        \makebox[.5\textwidth]{\raggedright\ignorespaces --\thepage{}--
        \hfill }}
\def\@evenhead{\@oddhead}
\def\subsectionmark##1{\markboth{##1}{}}
}

\ps@headings

\catcode`\@=12

\relax

\def\figcap{\section*{Figure Captions\markboth
        {FIGURECAPTIONS}{FIGURECAPTIONS}}\list
        {Fig. \arabic{enumi}:\hfill}{\settowidth\labelwidth{Fig. 999:}
        \leftmargin\labelwidth
        \advance\leftmargin\labelsep\usecounter{enumi}}}
 \relax
\def\tablecap{\section*{Table Captions\markboth
        {TABLECAPTIONS}{TABLECAPTIONS}}\list
        {Table \arabic{enumi}:\hfill}{\settowidth\labelwidth{Table 999:}
        \leftmargin\labelwidth
        \advance\leftmargin\labelsep\usecounter{enumi}}}
 \relax
\def\reflist{\section*{References\markboth
        {REFLIST}{REFLIST}}\list
        {[\arabic{enumi}]\hfill}{\settowidth\labelwidth{[999]}
        \leftmargin\labelwidth
        \advance\leftmargin\labelsep\usecounter{enumi}}}
 \relax

\catcode`\@=11

\def\marginnote#1{}

\def\ps@headings{\def\@oddfoot{}\def\@evenfoot{}
\def\@oddhead{\hbox{}\hfill
        \makebox[.5\textwidth]{\raggedright\ignorespaces --\thepage{}--
        \hfill }}
\def\@evenhead{\@oddhead}
\def\subsectionmark##1{\markboth{##1}{}}
}

\ps@headings

\relax

\def\firstpage#1#2#3#4#5#6{
\begin{document}
\begin{titlepage}
\nopagebreak
\title{\begin{flushright}
        \vspace*{-1.8in}
        {\normalsize CERN--TH/97-184}\\[-9mm]
        {\normalsize IC/97/93}\\[-9mm]
   {\normalsize CPTH--S552.0797}\\[-9mm]
       {\normalsize hep-th/9708075}\\[4mm]
\end{flushright}
{#3}}
\author{\large #4 \\ #5}
\maketitle
\vskip -7mm     
\nopagebreak 
\def\baselinestretch{1.0}
\begin{abstract}
{\noindent #6}
\end{abstract}
\vfill
\begin{flushleft}
\rule{16.1cm}{0.2mm}\\[-3mm]
$^{\star}${\small Research supported in part by\vspace{-4mm}
the National Science Foundation under grant
PHY--96--02074, \linebreak and in part by EEC
under the TMR contracts    
ERBFMRX-CT96-0090 and FMRX-CT96-0012.}\\[-3mm]
$^{\dagger}${\small Laboratoire Propre du CNRS UPR A.0014.}\\
CERN-TH/97-184\\
August 1997
\end{flushleft}
\thispagestyle{empty}
\end{titlepage}}
\newcommand{\Zint}{{\mbox{\sf Z\hspace{-3.2mm} Z}}}
\newcommand{\Real}{{\mbox{I\hspace{-2.2mm} R}}}
\def\simlt{\stackrel{<}{{}_\sim}}
\def\simgt{\stackrel{>}{{}_\sim}}
\newcommand{\dal}{\raisebox{0.085cm}
{\fbox{\rule{0cm}{0.07cm}\,}}}
\newcommand{\dt}{\partial_{\langle T\rangle}}
\newcommand{\dtbar}{\partial_{\langle\bar{T}\rangle}}
\newcommand{\al}{\alpha^{\prime}}
\newcommand{\mst}{M_{\scriptscriptstyle \!S}}
\newcommand{\mpl}{M_{\scriptscriptstyle \!P}}
\newcommand{\dv}{\int{\rm d}^4x\sqrt{g}}
\newcommand{\lv}{\left\langle}
\newcommand{\rv}{\right\rangle}
\newcommand{\ph}{\varphi}
\newcommand{\abar}{\bar{a}}
\newcommand{\sbar}{\,\bar{\! S}}
\newcommand{\xbar}{\,\bar{\! X}}
\newcommand{\fbar}{\,\bar{\! F}}
\newcommand{\zbar}{\bar{z}}
\newcommand{\dbar}{\,\bar{\!\partial}}
\newcommand{\tbar}{\bar{T}}
\newcommand{\taubar}{\bar{\tau}}
\newcommand{\ubar}{\bar{U}}
\newcommand{\tetabar}{\bar\Theta}
\newcommand{\etabar}{\bar\eta}
\newcommand{\qbar}{\bar q}
\newcommand{\ybar}{\bar{Y}}
\newcommand{\phb}{\bar{\varphi}}
\newcommand{\cm}{Commun.\ Math.\ Phys.~}
\newcommand{\prl}{Phys.\ Rev.\ Lett.~}
\newcommand{\pr}{Phys.\ Rev.\ D~}
\newcommand{\pl}{Phys.\ Lett.\ B~}
\newcommand{\ibar}{\bar{\imath}}
\newcommand{\jbar}{\bar{\jmath}}
\newcommand{\np}{Nucl.\ Phys.\ B~}
\newcommand{\F}{{\cal F}}
\renewcommand{\L}{{\cal L}}
\newcommand{\A}{{\cal A}}
\newcommand{\M}{{\cal M}}
\newcommand{\K}{{\cal K}}
\newcommand{\T}{{\cal T}}
\renewcommand{\Im}{\mbox{Im}}
\newcommand{\e}{{\rm e}}
\newcommand{\be}{\begin{equation}}
\newcommand{\en}{\end{equation}}
\newcommand{\gsi}{\,\raisebox{-0.13cm}{$\stackrel{\textstyle
>}{\textstyle\sim}$}\,}
\newcommand{\lsi}{\,\raisebox{-0.13cm}{$\stackrel{\textstyle
<}{\textstyle\sim}$}\,}
\date{}
\firstpage{????}{}
{\large\sc Duality in Superstring Compactifications\\[-5mm] 
with Magnetic Field Backgrounds$^\star$}
{I. Antoniadis$^{\,a,b}$, E. Gava$^{c,d,e}$, K.S. Narain$^{e}$ and
T.R. Taylor$^{\,f}$}
{\normalsize\sl $^a$Theory Division, CERN, 1211 Geneva 23,
Switzerland\\[-3mm]
$^b$\normalsize\sl
Centre de Physique Th\'eorique, Ecole Polytechnique,$^\dagger$
F-91128 Palaiseau, France\\[-3mm]
\normalsize\sl
$^c$Istituto Nazionale di Fisica Nucleare, sez.\ di Trieste,
Italy\\[-3mm]
\normalsize\sl
$^d$Scuola Internazionale Superiore 
di Studi Avanzati, Trieste, Italy\\[-3mm]
\normalsize\sl $^e$International Centre for Theoretical Physics,
I-34100 Trieste, Italy\\[-3mm]
\normalsize\sl $^f$Department of Physics, Northeastern
University, Boston, MA 02115, U.S.A.}
{Motivated by the work of Polchinski and Strominger on type IIA theory,
where the effect of non-trivial field strengths for $p$-form potentials 
on a Calabi--Yau space was discussed, we study four-dimensional heterotic 
string theory in the presence of a magnetic field
on a 2-cycle in the internal manifold, for both $N=4$ and $N=2$ cases. We
show that at special
points in the moduli space, certain perturbative charged states become 
tachyonic  and stabilize the vacuum by acquiring vacuum expectation
values, thereby  restoring supersymmetry. We discuss both the cases where
the tachyons appear with a tower of Landau levels, which become light in
the limit of large volume 
of the 2-cycle, and the case where such Landau levels are not present. In the
latter case it is sufficient to restrict the analysis to the quartic potential
for the tachyon. On the other hand, in the former case it is necessary to
include the
Landau levels in the analysis of the potential; for toroidal and orbifold
examples, we give an explicit CFT description of the new supersymmetric vacuum. 
The resulting new vacuum
turns out to be in the same class as the original
supersymmetric one.  
Finally, using duality, we discuss the role of the Landau levels on the
type IIA  side.}
\setcounter{section}{0}

\section {\bf Introduction}

In Ref.\ \cite{ps}, breaking of $N=2$ supersymmetry in four dimensions was
considered in type IIA theory as a consequence of non-trivial expectation
values for various field strengths on the Calabi--Yau space. By choosing
2-, 4-, 6- and 10-form field strengths one could gauge the translational
symmetry of the Neveu--Schwarz (NS) axion in all possible electric and
magnetic directions. 
This leads to a scalar potential that comes from the gauge kinetic term of
the ten-dimensional theory. 
With the exception of the 10-form field strength, for which
one has to start from the massive type IIA theory, the remaining cases,
for small field strengths, can be studied within the effective field
theory starting from the conventional type IIA in $D=10$ and  doing a
Kaluza--Klein reduction. We recall that in the sigma-model frame 
the type IIA bosonic action in $D=10$ is:
\begin{eqnarray}
S&=& \frac{1}{2}\int d^{10}x \bigg[\sqrt{-g}\bigg\{e^{-2\ph} 
\left[-R +4(\partial \ph)^2 -
\frac{3}{4}(dB)^2\right] \nonumber \\ &&\hskip 1.5cm +\; \frac{1}{4}F_2^2 + 
\frac{3}{4}(F_4-2dB\wedge A)^2\bigg\} + \frac{1}{64}F_4\wedge F_4 \wedge 
B\bigg], 
\label{s}
\end{eqnarray}
where $B$ is the
NS-NS antisymmetric tensor field, $\ph$ is the dilaton, and $F_2$ and $F_4$ are
the field strengths of the Ramond--Ramond (R-R) 1-form and 3-form potentials, 
respectively.

Consider for instance, the case where the 6-form field strength $F_6$ (the dual
of $F_4$) gets an expectation value $F_6 = \nu_6 \omega_6$, where $\omega_6$ is
the volume form of the Calabi--Yau space and $\nu_6$ is a constant that is
quantized in units of its inverse volume $V$. The ten-dimensional term
$F_6\wedge F_2\wedge B$ will then give the four-dimensional coupling 
$A_\mu \partial_\mu a$, $A_\mu$
being the graviphoton gauge potential and $a$ the axion field. Thus, an 
expectation value of $F_6$ leads to an electric gauging of the translation 
symmetry of the axion. The resulting
scalar potential, due to the kinetic energy of $F_4$,
is $g_{II}^4\nu_6^2 V\sim g_{II}^4/V$ (in the Einstein frame), 
where $g_{II}=e^\ph$ is
the four-dimensional type IIA coupling. The fact that the coupling appears
to the fourth power is due to the R-R nature of the gauge fields. This
potential is runaway and vanishes only in the limit of infinite volume or
vanishing coupling. However, it was argued that it can be stabilized in the
presence of additional massless charged states in the unbroken theory
\cite{ps}.  In the
type II theory, such massless states do indeed appear at the conifold points. 
In the presence of non-trivial field strengths these states become tachyonic 
and, after
minimizing the quartic potential with respect to these tachyonic fields, one 
finds a new $N=2$ supersymmetric vacuum. 

Our aim in this paper is to study the above phenomenon in the context of 
heterotic
theory compactified on $K_3 \times T^2$, where these tachyonic states appear 
at the perturbative level. As a result,
one can carry out the analysis in a more precise and complete way. We will see 
below
that in the presence of a magnetic field on a two-sphere in $K_3$, the
tachyon state is in general accompanied
by an infinite tower of excitations that correspond to higher Landau levels 
whose mass squares 
are of the order of the inverse power of the volume of the two-sphere. In the
context of $N=2$ theories these higher Landau levels are non-BPS states and
are therefore not stable. Nevertheless they contribute to the effective
potential with terms of higher powers of
the tachyon field that cannot be ignored in the large volume limit of the
two-sphere, since the contributions they give to the vacuum energy
are of the same order as the ones due to the quartic term. 
The large volume limit  is mapped, via duality, to the weak coupling limit of
IIA
theory, where these higher Landau levels appear as branes on non-supersymmetric
cycles whose masses are of the order of the coupling constant. The
minimization of the potential
therefore requires an analysis including all of these states. 

We find that supersymmetry is restored at the new minimum
and show that it is in the same class as the original
supersymmetric vacuum before turning on the magnetic field. We will also study 
the case where all the excitations have mass splittings of the order of
the string scale. In this case
it is sufficient to restrict the analysis to the quartic terms in the
tachyon effective potential, as higher powers in the tachyon field give, in
the large volume limit,
higher-order contributions to the vacuum energy.
 
The paper is organized as follows. In Section 2,
guided by the type II--heterotic
string duality, we show that  the heterotic version of the phenomenon
discussed in \cite{ps} involves turning on a magnetic field on
appropriate two-cycles of the $N=2$ compactification manifold, 
$K_3\times T^2$, or $T^6$ in the $N=4$ case. 
For simplicity, we restrict ourselves
to toroidal or orbifold compactifications. We also
discuss the spectrum of states in
the presence of a magnetic field. The tachyons that appear in the untwisted 
sector are accompanied by higher Landau levels, while those appearing in the
twisted sector are not. In Section 3, we discuss the examples involving higher
Landau levels both in $N=4$ and in $N=2$ theories. In the latter case, we
discuss in some detail examples of $\Zint_2$ and $\Zint_4$ orbifold models. 
In Section 4, we study the case of twisted
sector tachyons that do not appear with Landau levels. Section 5 contains a
discussion of the implication of these results for the 
type IIA side.
    
\section {\bf Heterotic Theory in the Presence of Magnetic Fields}

To anticipate what may happen in the heterotic string theory it is convenient
to start from the six-dimensional (6D)
type II--heterotic dual pair \cite{ht, wit}. 
Type IIA is compactified
on $K_3$ while heterotic on $T^4$. The 6D gauge fields in type IIA are
associated with the cohomology classes of $K_3$, namely one each from the 0- and
4-cohomology and 22 from the 2-cohomology. On the heterotic side these gauge
fields appear in the usual way: 4 from the left-moving fermionic sector and 20
from the right-moving bosonic sector. At a generic point in the moduli space,
the gauge group is Abelian. Now let us further compactify the 6D
theory on a two-dimensional compact space, which can be either $T^2$ or
$CP^1$. In the
$T^2$ case the resulting 4D theory has $N=4$ supersymmetry, while in the
$CP^1$ case, in order to have conformal invariance, one must consider a
fibration of the 6D theory over the base $CP^1$ and the resulting 4D theory
has $N=2$ supersymmetry \cite{kv,fhsv,klm}. In the type IIA case
this gives rise to a Calabi--Yau space,
which is a $K_3$ fibration over the base, while in the heterotic case, this can
be thought of as a compactification on $K_3\times T^2$, together with an
appropriate gauge bundle. The base $CP^1$ in this case is part of $K_3$.
Under the duality, the 4D inverse coupling squares of type II
($g^{-2}_{II}$) and heterotic ($g^{-2}_{H}$) are mapped to the volume of
the base of the heterotic ($V_{H}$) and type II ($V_{II}$), respectively.  

Thus, if one wants to describe the above phenomenon of supersymmetry
restoration
perturbatively on the heterotic side, one should consider those gaugings on
the type IIA side for which the scalar potential vanishes in the limit of large
volume of the base. If we give expectation values to the field strengths of the
6D gauge potentials (that survive the fibration) on the base, then these
field strengths are quantized via the Dirac quantization condition. In the limit
of large $V_{II}$, where the metric of the Calabi--Yau factorizes into a
component
along the fibre and one along the base, one can see that the field strength is
quantized in units of the inverse volume $V_{II}$ of the base. 
The kinetic term for the gauge field then gives rise to a 4D tree-level 
potential. As we mentioned above, in the large $V_{II}$ limit, the
resulting potential  (in the
Einstein frame) is of order $g_{II}^4/V_{II}$. From the
$D=10$ point of view, these gauge field strengths correspond to 2-, 4- and
6-forms that have non-vanishing integrals over the base. 
On the heterotic side, such field
strengths will have the interpretation of 6D gauge fields,
acquiring non-vanishing field strengths on the base. 

On the heterotic side there is a 6D tree-level coupling $dB\cdot\omega(A)$, with
$B$ being the antisymmetric tensor and $\omega$ the Chern--Simons form. 
It follows
that a non-trivial field strength on the base induces in four dimensions a term
$A_\mu\partial_\mu b$, where $b$ is the pseudoscalar modulus associated to the
K\"ahler class of the base. This mechanism therefore gauges the translation
symmetry of $b$. This is expected from the fact that the type IIA axion is
mapped to $b$ under duality.\footnote{This translation symmetry is of course
broken by world-sheet instanton effects on the heterotic side, in the same way
as the axion shift is broken by space-time non-perturbative effects on the
type II side.} Note that in the heterotic side this gauging is always electric.
 
By the usual Dirac quantization condition, the expectation value of the field
strength is quantized in units of the inverse volume $V_{H}$ of the base, in
the limit of large $V_H$. 
Note that under duality the modulus $V_{H}$ is mapped to the type II dilaton 
and belongs to a hypermultiplet in the case of $N=2$ compactifications.
The 6D gauge kinetic term then  gives rise to a tree-level cosmological 
constant that falls off as $g_H^2/V_H^2$ (in the Einstein frame).
Taking into account the relation between coupling and volume in the two 
theories, we see that their scalar potentials match one another.
As discussed in Ref.\ \cite{ps}, this runaway potential can be stabilized
if there are special points in the vector moduli space at which there
appear extra massless states, which are charged under the above gauge
field. In heterotic perturbation theory, such states can arise only for
gauge fields coming from the right-moving sector. In the following we will
restrict ourselves to such gaugings. 

In Refs.\ \cite{no,b,kk}, the spectrum in the presence of a constant
magnetic field $F$  for some $U(1)$ gauge field on a torus of volume $V$
has been analysed.  Given the minimum charge 
$q_{\rm min}$, the single-valuedness of the wave function of charged states
implies the following quantization for the magnetic field:
\begin{equation}
F={2\pi k\over q_{\rm min}V}\ ,
\label{quant}
\end{equation}
for some integer $k$. Then, the previously massless states, with charge
$q=\ell q_{\rm min}$ in
$D=6$,  give rise to a tower of Landau levels with masses (for small $F$):
\begin{equation}
M^2=g_H^2\left[ (2n+1)|qF|-2sqF\right]\ ,
\label{levels}
\end{equation}
and multiplicity $\ell k$. Here,
$s$ is the eigenvalue of the internal spin operator,
and the non-negative integer $n$
labels the Landau levels. Landau levels arise from the quantization
of the two compact momenta that do not commute in the presence
of a magnetic field.
In the toroidal (or orbifold) case there is always a
state (untwisted) with $s=1$; for $n=0$, this state becomes tachyonic, 
with mass squared
$-g_H^2|qF|$. Note that for fermions $s=\pm 1/2$, the two signs 
corresponding to the left and right
chirality in $D=4$, respectively. Thus one chirality (say left) fermion
(for $n=0)$ remains massless while the other chirality (right) becomes massive
and is paired with the left part of $n=1$ fermion and so on. These results,
obtained from field theory considerations, 
can also be verified in the context of
type I string theory, and the tree-level mass formula 
has been obtained to all orders in $F$ in Refs.\ \cite{b,acny}.

In the case of orbifold compactifications, there is also another class of
massless charged states, hypermultiplets 
arising from the twisted sector, which have no
six-dimensional interpretation. 
In the presence of a magnetic field, these states
can also become tachyonic; however, they are not accompanied by Landau levels
with mass splittings proportional to $F$ because they have no momenta in the
$K_3$ direction. This is most easily seen by going to 
the type I side \cite{bs,gp,dp}. Here the gauge group arises from 9- and 5-brane
Chan--Paton
charges, with the 9-brane gauge group corresponding to the heterotic
perturbative gauge group. The 5-branes are  located at fixed points
of the $K_3$ orbifold. The heterotic twisted orbifold states
are mapped to ground states of open strings stretched between 9- and 5-branes.
Note that these strings have mixed Neumann--Dirichlet (ND) boundary conditions;
therefore they do not carry Kaluza--Klein $K_3$ momentum, similarly
to the heterotic case. The magnetic field is now on a 9-brane plane,
orthogonal to the 5-branes. The effect of such a field on 9-brane
is to modify the boundary condition of the open string on the 9-brane end from
Neumann to a generalized boundary condition. Therefore the 95 string, that
involved half-integer oscillator modes in the ND directions, now involves
oscillators with frequencies $(1/2 \pm qF)$ modulo integers (in the weak field
approximation). As a result, the zero point energy gets shifted and 
the previously massless scalars split into two states $\chi$, $\eta$, with
masses: 
\begin{equation}
M^2_{\chi}=-|qF|,~~~~~~ M^2_{\eta}=|qF|\ ,
\label{levelI}
\end{equation}
in the type I string frame, while the fermions remain massless.
Since the lowest oscillator mode has a frequency $(1/2-|qF|)$, it is clear that
the excited states have mass splittings of order the string scale and as a
result one does not have the usual Landau levels. 

As we will see below, since the presence of Landau levels affects 
the analysis of the potential, we will consider the cases
with and without them separately.

\section {\bf Tachyons with Landau Levels}

This case includes $N=4$ toroidal compactification, as
well as untwisted sector in $N=2$ orbifold compactifications 
of the heterotic string. Explicitly, in the untwisted sector, the tachyonic
modes corresponding 
to $s=\pm 1$ in Eq.\ (\ref{levels}) are [$(\partial X_4 \pm i
\partial X_5)e^{iP_L\cdot X_L + iP_R\cdot X_R}$ + the supersymmetric
completion], where $X_4, X_5$ are the coordinates of
the torus, and $P_L$ and $P_R$ are charges of the gauge fields coming from 
left- and 
right-moving sectors, respectively. These charges satisfy the condition 
$P_R^2 - P_L^2 = 2$. The squared masses of these states are $\frac{1}{2}P_L^2$. 
In the presence of the constant gauge field strength $F_{45}^a$ in some
right-moving direction labelled by $a$, these masses are shifted according to 
Eq.\ (\ref{levels}), 
with $q= P_R^a/\sqrt{2}$. By adjusting the Wilson 
lines, one can go to a point in the moduli space where $P_L=0$ 
and then the mass is just given by Eq.\ (\ref{levels}). 
In fact there are at least two states (or an even number) with the same mass, 
and they correspond to $P_R \rightarrow -P_R$ and $s 
\rightarrow -s$. As a result the massless fermions are paired and become 
non-chiral. This was to be expected, because by turning on the Wilson lines the
fermion can pick up a mass. Note that in this way we can never get a situation 
like the conifold, where only one hypermultiplet becomes massless.

The existence of tachyons in the spectrum signals instability; dynamically the
tachyon will start getting an expectation value. The question we would like to
address is whether there exists a new critical point where the potential
vanishes and the supersymmetry is restored. In order to get a feeling for what
might happen, let us consider the effective potential of the tachyon up to the
quartic term. To leading order in $1/V$, this quartic
potential arises from the Kaluza--Klein reduction of a 6D gauge theory,  
in the presence of a background magnetic field on the base. Since the
tachyons in question involve internal spins $s=\pm 1$, the
relevant gauge theory is that of $SU(2)$,
where the internal components of $W^\pm$ with internal helicities $s=\pm 1$
play the role of the tachyon. In the following we first discuss the $N=4$ case,
where the base manifold is $T^2$, and then discuss the modifications for the
$N=2$ case, when the base is realized as an orbifold.

\subsection{\large $N=4$ Case}

The discussion in this case is very similar to the one appearing in the 
bound-state problem of $p$- and $(p+2)$-branes \cite{gns}.
Since modular invariance implies that the complete spectrum must also include 
fundamental representations of $SU(2)$, the minimum value of the
background magnetic field is
$\frac{2\pi}{V}\sigma_3$, with $\sigma_3$ the Pauli matrix. 
This can be accomplished for instance by
choosing the background gauge potential to be $A^0_5 = 
\frac{2\pi}{V} \frac{x_4}{R_4}\sigma_3$ and all the remaining $A$'s equal to
zero. Note that this background field satisfies the periodicity conditions:
\begin{eqnarray}
A^0(x_4, x_5 +R_5) &=& \Omega_1 A^0(x_4,x_5), ~~~~ \Omega_1 = 1\nonumber \\
A^0(x_4+R_4,x_5) &=& \Omega_2 A^0(x_4,x_5), ~~~~ \Omega_2 = e^{2i\pi x_5
\sigma_3/R_5}\ ,
\label{bc}
\end{eqnarray}
where the symbol $\Omega A$ means the gauge transformation of $A$ by $\Omega$.
Note that this boundary condition corresponds to a trivial $\Zint_2$ flux:
\begin{equation}
\Omega_1(0,0) \Omega_2(0,R_5)= \Omega_2(0,0)\Omega_1(R_4,0)\ .
\label{toronbc}
\end{equation}
This fact will play an important role later. 

Let us parametrize the fluctuations of $A_4$ and $A_5$ (i.e.\ the scalars
in 4D space-time) around this background in the
following way:
\begin{equation}
A_z = A^0_z + \phi \sigma_3 + \chi \sigma_+ + \bar{\eta} \sigma_-, ~~~~~
A_{\bar{z}} = (A_z)^{\dagger}\ ,
\label{fluct}
\end{equation}
where $z=x_5 +ix_4$ and $\sigma_{\pm}= \sigma_1 \pm i\sigma_2$.
The boundary conditions on $\phi$, $\chi$ and $\eta$ 
are determined from the fact that $A$ in Eq.\ (\ref{fluct}) satisfies the 
same boundary conditions as
the background field, namely Eq.\ (\ref{bc}). This means that under
$x_5\rightarrow x_5+R_5$ they remain unchanged while under $x_4\rightarrow
x_4+R_4$ they transform as:
\begin{equation}
\phi \rightarrow \phi,~~~~\chi\rightarrow e^{4\pi i x_5/R_5} \chi,
~~~~\eta\rightarrow e^{4\pi i x_5/R_5}\eta\ .
\label{bcchi}
\end{equation}

To leading order in the large volume limit, the potential energy 
is given by $\int_{T^2} {\rm tr} F^2$, where $F$ is the field
strength along the $T^2$ directions. Explicitly, the components of $F$ are:
\begin{eqnarray}
F_3 &=& \frac{2\pi}{V}-i(\partial_{\bar{z}}\phi-\partial_z \bar{\phi}) + 
|\eta|^2 - |\chi|^2
\nonumber\\
F_+ &=& D_z\chi + D_z^{\dagger}\eta +2i\bar{\phi}\chi - 2i\phi \eta\ ,
\label{Fs}
\end{eqnarray}
where the subscripts in $F$ are the gauge indices, 
$D_z = \partial_z -2ix_4/V$, and $D_z^{\dagger}= -\partial_{\bar{z}}+2ix_4/V$
is the adjoint of $D_z$. From the analogy with the harmonic oscillator problem,
it is clear that $D_z$ acts as an annihilation operator while its adjoint
$D_z^{\dagger}$ acts as the creation operator. 
The ground state that is annihilated 
by $D_z$ and subject to the boundary condition (\ref{bcchi}) 
gives the two degenerate tachyon wave functions:
\begin{equation}
\Psi_0^{(1)}=Ne^{-\pi x_4^2/V}\theta_2(\tau|\bar{z})\quad ;\quad 
\Psi_0^{(2)}=Ne^{-\pi x_4^2/V}\theta_3(\tau|\bar{z})\quad ;\quad 
\tau\equiv 2i{R_4\over R_5}\ ,
\label{twf}
\end{equation}
where $\theta$'s are the Jacobi theta-functions and
$N=(-2i\pi\tau)^{1/4}$ is a normalization factor such that
$\int_{T^2}|\Psi_0^{(i)}|^2=V$ for $i=1,2$. 
Their mass is given by Eq.\ (\ref{levels}) for $n=0$ and
$s=1$. The wave functions for higher Landau levels are simply obtained by
applying the creation operators $D_z^{\dagger}$ on $\Psi_0^{(i)}$ 
and using the background gauge condition $D_z\chi = D_z^{\dagger}\eta$.

The $\chi$ field can then be expanded in terms of orthonormal wave functions
$\Psi_n^{(i)}$'s obtained by applying the creation operators 
${D_z^{\dagger}}^n$ on $\Psi_0^{(i)}$. 
Thus, $\chi=\sum_n \chi_n^{(i)}\Psi_n^{(i)}$, with some complex coefficients 
$\chi_n^{(i)}$ that play the role of four-dimensional scalar fields. 
Similarly, one can expand the field $\eta=\sum_n \eta_n^{(i)}\Psi_n^{(i)}$.
The background gauge condition can be used to solve $\eta$ in terms of $\chi$
with the result $\eta_n^{(i)}=\sqrt{n+2\over n+1}\chi_{n+2}^{(i)}$, 
while one also finds that
$\chi_1^{(i)}=0$. Substituting these expansions in the action and rescaling
$\chi_{n+2}^{(i)}\rightarrow\sqrt{n+1\over 2n+3}\chi_{n+2}^{(i)}$ 
in order to get standard
kinetic terms, we find that their masses are $(2n+3)g_H^2/V$, in agreement with
the mass fomula (\ref{levels}). In fact, Eq.\ (\ref{levels}) with $s=\pm 1$ 
implies that these states come with multiplicity 4 while the tachyon and the 
first excitation with mass $g_H^2/V$ are twofold-degenerate. 
On the other hand, the
transverse vectors corresponding to $s=0$ are twofold-degenerate 
and have masses $(2n+1)g_H^2/V$.
Combining all these states into massive Lorentz representations, one finds that
there are two massive spin 1 representations for each level $(2n+1)g_H^2/V$ and
two scalars for each level $(2n+3)g_H^2/V$, besides the tachyons. These scalars
are precisely the ones we have found above by mode expansion and 
which enter in the minimization of the scalar potential.

The background gauge condition for the field $\phi$ is 
$\partial_{\bar{z}}\phi+\partial_z \bar{\phi}=0$. Since $\phi$ is periodic on
the torus as seen from Eq.\ (\ref{bcchi}), its mode expansion subject to
the gauge condition is given by: 
\begin{equation}
\phi=\phi_0+ i\sum_{p\neq 0}{\rm arg}(p) \phi_p e^{2i\pi p\cdot x}\quad ;\quad
\phi_{-p}=\bar{\phi}_p\ ,
\label{phip}
\end{equation}
with momenta $p\equiv (p_4,p_5)=(m_4/R_4, m_5/R_5)$ for integers $m_{4,5}$ and
${\rm arg}(p)={\rm arctan}(p_4/p_5)$. 
The mass of the mode $\phi_p$ is $g_H^2|p|$.

The tachyon effective potential up to the quartic level has an irreducible
part, which can be read off from $\int_{T^2}{\rm tr} F^2$:
\begin{eqnarray}
{\cal{V}}_{\rm \rm eff}^{\rm irr}&=&
\frac{1}{2} g_H^2 \bigg\{\left({2\pi\over V}\right)^2 - 
{4\pi\over V}(|\chi_0^{(1)}|^2 + |\chi_0^{(2)}|^2)
\; + \;\theta_3(\tau)\theta_3(-1/\tau)\,(|\chi_0^{(1)}|^4+|\chi_0^{(2)}|^4)
\nonumber\\ 
&& +~4\theta_3(\tau)\theta_4(-1/\tau)\,|\chi_0^{(1)}|^2|\chi_0^{(2)}|^2\;+\;
2\theta_2(\tau)\theta_4(-1/\tau)\,[(\chi_0^{(1)}{\bar\chi}_0^{(2)})^2+
{\rm c.c.}] 
\bigg\}\ ,
\label{potir}
\end{eqnarray}
where the appearance of $\theta$-functions is a result of integration of the
four-tachyon wave functions (\ref{twf}). 
In addition to these irreducible terms, 
the effective potential also
receives reducible contributions due to the exchange of the Kaluza--Klein modes
$\phi_p$'s through the following 3-point vertex:
\begin{equation}
C_p^{(i,j)} \phi_p\chi_0^{(i)}{\bar\chi}_0^{(j)}\quad ,\quad 
C_p^{(i,j)}=|p|e^{-i\pi m_4({m_5\over 2}-j)}
e^{i{\pi\over 2}({p_5^2\over 4}\tau-{p_4^2\over\tau})}
\quad ;\quad m_5+i-j={\rm even}\ ,
\label{cp}
\end{equation}
where the structure constants $C_p^{(i,j)}$ arise upon integration 
of the appropriate wave functions, and they vanish if $p_5$ does not satisfy the
above condition.
It is easy to check that the contribution of the reducible 
diagrams is:
\begin{eqnarray}
{\cal{V}}_{\rm eff}^{\rm red}&=&\frac{1}{2} g_H^2 \bigg\{
[1-\theta_3(\tau)\theta_3(-1/\tau)]\,(|\chi_0^{(1)}|^4+|\chi_0^{(2)}|^4)
+ [2-4\theta_3(\tau)\theta_4(-1/\tau)]\,|\chi_0^{(1)}|^2|\chi_0^{(2)}|^2
\nonumber\\ 
&& -~2\theta_2(\tau)\theta_4(-1/\tau)\,[(\chi_0^{(1)}{\bar\chi}_0^{(2)})^2
+{\rm c.c.}]
\bigg\}\ ,
\label{potred}
\end{eqnarray}
As a result, the
effective potential up to this order becomes a total square:
\begin{equation}
{\cal{V}}_{\rm eff}=\frac{1}{2} g_H^2 \;
\bigg({2\pi\over V}-|\chi_0^{(1)}|^2-|\chi_0^{(2)}|^2\bigg)^2 \ .
\label{pot}
\end{equation}
Now it is clear that the minimum of the potential is at 
$\sum_i|\chi_0^{(i)}|^2=2\pi/V$ and
that at this point the potential vanishes.

Let us now consider the higher-order terms in the tachyon effective potential.
These terms can either come from higher-derivative terms in $D=6$ (e.g.\ $F^4$
terms) or from reducible diagrams of higher-point functions. It is easy to see
that all the contributions arising from the higher-derivative interactions are
suppressed in the large volume limit. However, the reducible diagrams involving
vertices coming from $F^2$ lead to contributions that are of the same order
in volume as the quartic potential considered above. Indeed, the $n$-point
vertex scales as $V^{n-4 \over 2}$, while 
the propagators of the massive fields scale as 
$V$. A simple counting then shows that $|\chi_0|^{2n}$ scales as $V^{n-2}$. The
fact that the reducible diagrams involving exchanges of massive fields
$\phi_p$'s and the higher Landau levels $\chi_n$'s contribute just means that
these fields also acquire expectation values. It is therefore more efficient to
analyse the minimization problem directly at the six-dimensional level. 

Since the
potential $\int_{T^2} (F_3^2 + |F_+|^2)$ is a sum of integrals of semi-positive 
definite functions, the only way the potential can vanish is if
each function vanishes individually.  This means that the field strengths $F$'s
must vanish. The question therefore is whether one can continuously change the
$SU(2)$ field strength from the original non-zero value to a vanishing one
respecting the boundary conditions (\ref{bc}) and (\ref{toronbc}). 
This has been shown to
be possible in Refs.\ \cite{af,gth} for general toron boundary conditions.
In fact, our case (\ref{toronbc}) corresponds to the trivial $\Zint_2$
flux. By a suitable $SU(2)$ gauge transformation one can set the boundary
conditions $\Omega_1$ and  $\Omega_2$ to be constant commuting group
elements. The effect of these constant commuting elements consists in
turning on appropriate Wilson lines. As a result the new vacuum belongs to
the same class as the one of the original supersymmetric theory, and the
effect of the magnetic field is completely washed out. 

\subsection{\large $N=2$ Case}
 
Although the above discussion was made for the case of toroidal
compactification (i.e. $N=4$ theory), it can be extended to the case of
the untwisted sector in orbifold compactifications that give $N=2$ theories,
provided the tachyons are not projected out by the orbifold group.
To be concrete, we consider here the examples of $\Zint_2$  and 
$\Zint_4$ orbifolds \cite{wal}. In the former case the would-be tachyons
appear in pairs, which gives an
extra flat direction in the supersymmetric theory, while in the latter
case there is only one charged massless hypermultiplet at the special point in
the moduli space, which provides a tachyonic mode in the presence of a 
magnetic field. As a result there is no flat direction associated to it.

\noindent {\bf $\Zint_2$ orbifold}

Consider the theory in $D=4$ obtained by compactifying the ten-dimensional
theory on $(T^2 \times T^2/\Zint_2) \times T^2$ together with a shift
$\delta$ in one of the
$E_8$'s subject to the condition $2\delta \in \Gamma_8$, where $\Gamma_8$ is the
$E_8$ lattice, and the level matching condition $\delta^2 = {1 \over 2}$
mod
$\Zint$. If one wishes to go to $D=6$, one can decompactify the last
$T^2$
to get an $N=1$ theory. The six-dimensional anomaly cancellation provides
a further test on the analysis of the vacuum given below.

Let us now turn on a field strength on one of the $\Zint_2$-twisted tori
(say $x_4$ and $x_5$ directions of radii $R_4$ and $R_5$). Decomposing
$E_8$ in terms of  $E_7 \times SU(2)$,
let us take the field strength to be  along the Cartan direction of
the $SU(2)$. Depending on the choice of the orbifold shift $\delta$, this
$SU(2)$ may or may not be broken. In the case when $SU(2)$ is not broken,
we have to go to
lower dimensions and go to the Coulomb phase where $SU(2)$ is broken down
to $U(1)$, in order to discuss the quantization 
condition for the magnetic flux.
Let us choose a fundamental domain for $T^2/\Zint_2$ as $0 \leq x_4 \leq
R_4/2$ and $-R_5/4 \leq x_5 \leq 3R_5/4$ with the identifications on the
boundary given by the
orbifold group action: the boundary $x_5=-R_5/4$ is identified with
$x_5=3R_5/4$ and, at the boundaries $x_4=0$ and $x_4=R_4/2$, $x_5$ is identified
with $-x_5$ for $|x_5| \leq R_5/4$ and with $R_5-x_5$ for $|x_5| \geq
R_5/2$ (see Fig.\ 1). The
fixed points of the orbifold
are at $(x_4,x_5)$ given by $(0,0)$, $(0,R_5/2)$, $(R_4/2,0)$ and 
$(R_4/2, R_5/2)$.
\begin{figure}
\[
\psannotate{\psboxto(0cm;5cm){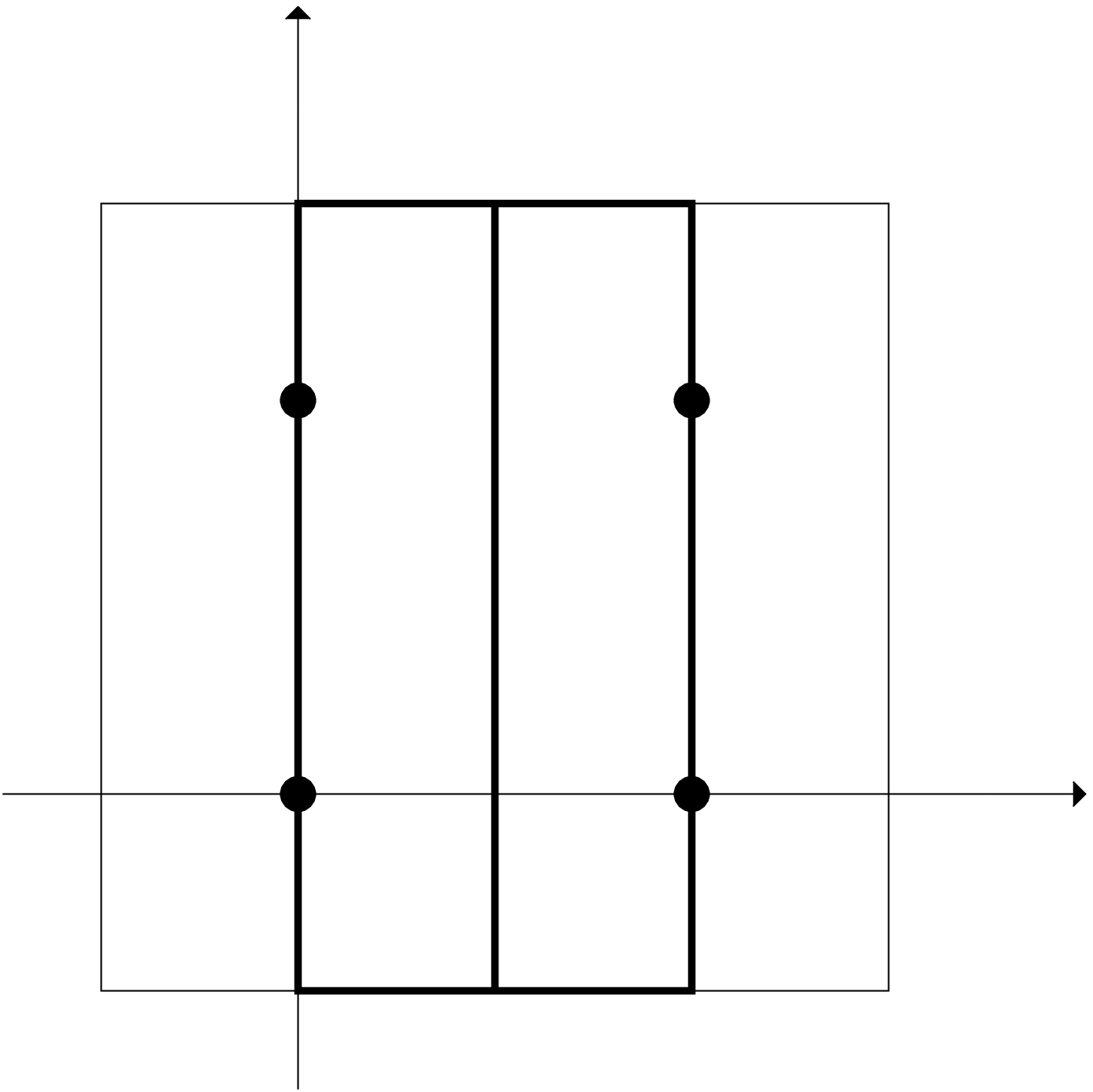}}
{
\at{10.5\pscm}{-1.8\pscm}{$ \Zint_2$}
\at{6.6\pscm}{13.8\pscm}{$x_5$}
\at{8\pscm}{12\pscm}{$\scriptstyle\frac{3}{4}R_5$}
\at{7.7\pscm}{2.7\pscm}{$0$}
\at{7.3\pscm}{8.6\pscm}{$\scriptstyle\frac{1}{2}R_5$}
\at{6.7\pscm}{0.2\pscm}{$\scriptstyle -\frac{1}{4}R_5$}
\at{7.1\pscm}{6.3\pscm}{\bf I}
\at{9\pscm}{6.3\pscm}{\bf II}
\at{15.2\pscm}{2.9\pscm}{$x_4$}
\at{12.8\pscm}{4.5\pscm}{$\scriptstyle\frac{3}{4}R_4$}
\at{9.8\pscm}{4.5\pscm}{$\scriptstyle\frac{1}{2}R_4$}
\at{-0.8\pscm}{4.5\pscm}{$\scriptstyle -\frac{1}{4}R_4$}
}
\hskip 1.5in
\psannotate{\psboxto(0cm;5cm){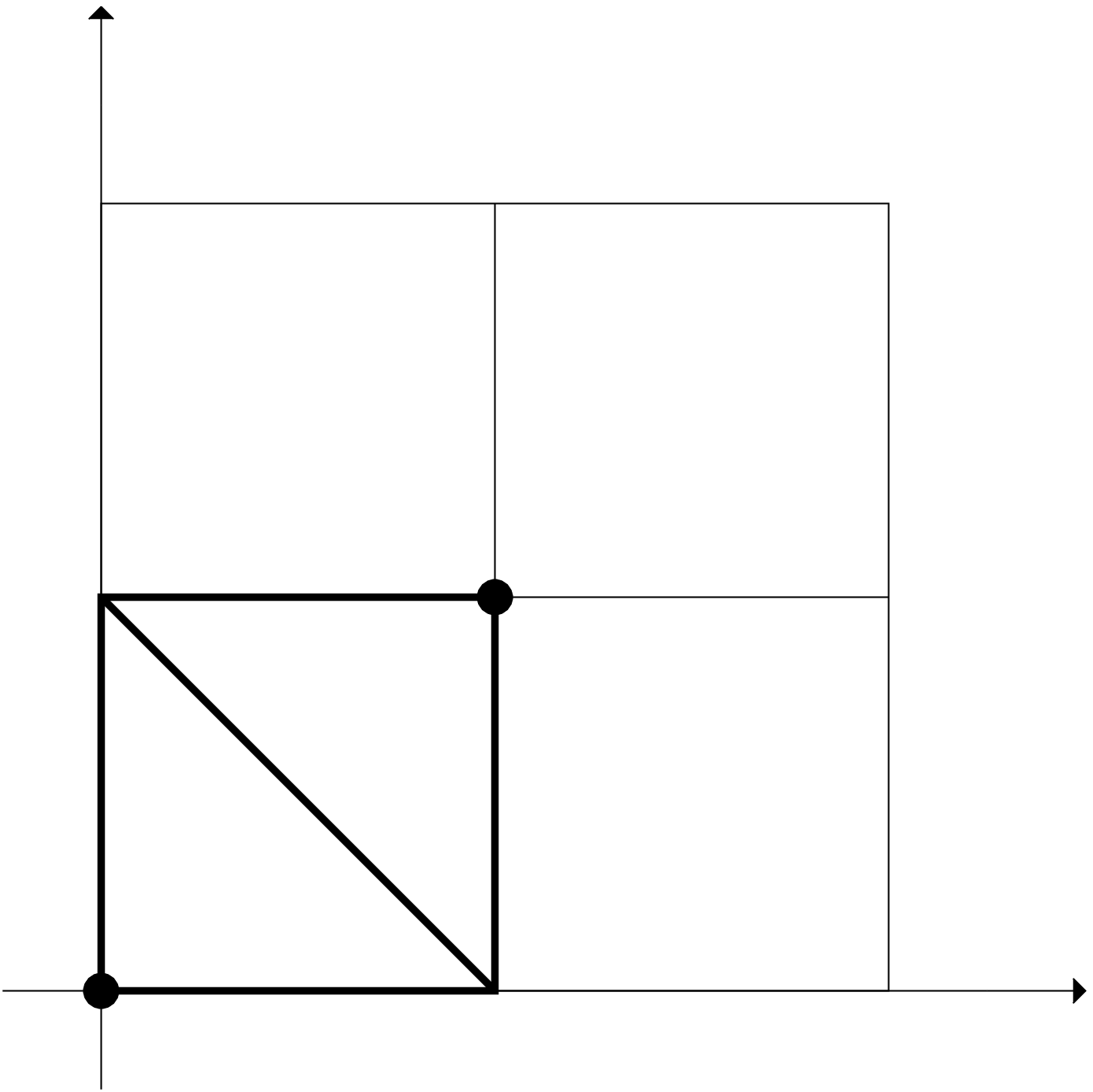}}
{
\at{3.7\pscm}{13.7\pscm}{$x_5$}
\at{5\pscm}{12\pscm}{$\scriptstyle R$}
\at{4.8\pscm}{0.2\pscm}{$0$}
\at{5.4\pscm}{2.6\pscm}{\bf I}
\at{6.6\pscm}{4.3\pscm}{\bf II}
\at{8.5\pscm}{1.8\pscm}{$\scriptstyle\frac{1}{2}R$}
\at{3.1\pscm}{7\pscm}{$\scriptstyle\frac{1}{2}R$}
\at{12.9\pscm}{1.8\pscm}{$\scriptstyle R$}
\at{14.4\pscm}{0.3\pscm}{$x_4$}
\at{6.3\pscm}{-1.8\pscm}{$ \Zint_4$}
}
\]
\caption{\it The fundamental domains, $T^2/ \Zint_2$
and $T^2/\Zint_4$, for $\Zint_2$ and $\Zint_4$ orbifolds, respectively.
The two patches, I and II, are defined in the text
and the black dots are the fixed points.
\label{domains}}
\end{figure}

In order to define the gauge potential associated with the $U(1)$ magnetic 
flux, we
choose two patches in the fundamental domain: patch I is given by 
$0 \leq x_4 \leq
R_4/4$ and patch II by $R_4/4 \leq x_4 \leq R_4/2$ (see Fig.\ 1).
We can now choose a gauge such that the $U(1)$ gauge potential $A_4=0$ and
$A_5 = B x_4 Q$ in patch I and $A_5 = B(x_4- R_4/2)Q$ in patch II, where $Q$ 
is the $U(1)$ generator. This choice ensures
that, at the boundaries, the gauge potential is well defined under the orbifold
group identification mentioned above. The gauge transition function $g_{II,I}$ 
in the overlap between the two patches, i.e.\ at $x_4 = R_4/4$, is
$g_{II,I}= \exp (-iB QR_4x_5 /2)$. Demanding single valuedness of this
transition function under $x_5 \rightarrow x_5 + R_5$ one
obtains the quantization condition $B= {{4\pi n} 
\over {q_{\rm min} R_4 R_5}}$, 
where $q_{\rm min}$ is the minimum charge and $n$ is an integer. Note that this 
quantization condition is twice that on the torus and is due to the 
fact that the area of the
orbifold $T^2/\Zint_2$ is half that of the torus $T^2$. Taking into
account that in our case this $U(1)$ is embedded in an $SU(2)$ and that
the charge spectrum also includes the fundamental
representation of $SU(2)$ we have that $Q=\sigma_3$ and $q_{\rm min}=1$. Thus
\begin{equation}
g_{II,I} = \exp\left(-2i\pi n {x_5 \over {R_5}} \sigma_3\right)\ .
\label{g21}
\end{equation}

Let us now go to the $SU(2)$ point by suitably adjusting the scalars of the 
vector multiplets. We would like to show that one can perform $SU(2)$ gauge
transformations that respect the orbifold group identifications, such that the
transition function $g_{II,I}$ becomes constant. In general if one performs a 
gauge transformation by $g_I$ in patch I and by $g_{II}$ in patch II, then the 
new transition function becomes $g'_{II,I} = g_{II} g_{II,I} g_I^{-1}$. Let us
make the ansatz $g_{II}=1$ and
\begin{equation}
g_I = h(x_4) e^{-i\pi n {x_5 \over {R_5}}\sigma_3} e^{if(x_4)\sigma_1}  
e^{i\pi n {x_5 \over {R_5}}\sigma_3} e^{-if(x_4)\sigma_1}\ ,  
\label{gI}
\end{equation}
where $f$ and $h$ are smooth and  $f(0) =0$, $f(R_4/4) = \pi/2$ and the $SU(2)$
group element $h$ satisfies
the condition $\Omega h(0)
=h(0)\Omega$, where $\Omega$ is the orbifold group action corresponding to
the shift $\delta$. These conditions ensure that $g_I$ is well defined on
the orbifold. The new transition function then becomes 
\begin{equation}
g'_{II,I} = h^{-1}(R_4/4) \equiv h^{-1}_0\ .
\label{g'}
\end{equation}
Since $g'_{II,I}$ is a constant, it follows that one can smoothly change the
background gauge field strength to zero and as a result one again obtains a 
supersymmetric vacuum. 

We would like to make a couple of remarks here. 
First, for a gauge transformation
$g_I$ to exist on patch I such that the new transition function $g'_{II,I}$
is a constant, the quantization condition (\ref{g21}), which ensures that
$g_{II,I}$ maps the boundary of patch I into a closed contour on the
$SU(2)$ group manifold, is crucial. If, for instance, one considers 
a theory with no fundamental representations of $SU(2)$, then the
quantization condition
would imply that $n$ in (\ref{g21}) could also be a half-integer. For $n$ 
half-integer, $g_{II,I}$ maps the boundary of patch I into an open path in the
$SU(2)$ group manifold whose end-points are related by the action of the 
non-trivial element of the centre $\Zint_2$. Since the patches are
topologically discs (due to orbifold identifications), it would be impossible
to find a smooth $g_I$ such that the new transition function becomes a constant.
Consequently the field strength would have been non-zero, giving rise to
a cosmological constant.
It is quite curious that in string theory, modular invariance guarantees that
the spectrum contains also
fundamental representations. In turn, this
enforces $n$ to be an integer and, as a result, the existence of a new 
supersymmetric
minimum is guaranteed. For the type II theories that are dual to the
heterotic string vacua, one can use the perturbative modular invariance
on the heterotic side to deduce the charge spectrum of the non-perturbative
states in the type II side. It will be interesting to find out whether there
is some generalized notion of non-perturbative modular invariance that
would impose conditions on the charge spectrum for non-perturbative
states for a generic type II theory (that may not have a heterotic dual). 
This is clearly important, as the existence of the new supersymmetric minimum
depends on the quantization condition, and hence on the charge spectrum.

The second remark concerns the role of the tachyon in the above solution 
(\ref{g'}). In order to arrive at the constant transition function, 
it is necessary that the gauge transformation $g_I$ involves also the 
off-diagonal elements of $SU(2)$.
This is so because, for non-zero $n$, $g_{II,I}$ maps the boundary of patch I 
into a closed path in the $U(1)$ subgroup (generated by $\sigma_3$)
that has winding number $n$. This path is non-trivial if one restricts oneself
to this $U(1)$ subgroup
and becomes trivial only in the $SU(2)$ group manifold; 
hence $g_I$ must necessarily
involve off-diagonal elements. Transforming now the original $U(1)$ 
gauge potential
by $g_I$ will necessarily introduce an off-diagonal part in the gauge 
potential; because of
the orbifold group invariance of $g_I$, this corresponds to turning
on expectation values for the tachyon fields (and all the higher Landau levels).
The final solution $A=0$ then comes about by a complicated field redefinition
using the orbifold group invariant part of the local
$SU(2)$ symmetry in six-dimensional space-time.

Just as in the $N=4$ case, the role of the constant group element $h_0$
is to turn on Wilson lines. To see this, let us extend the fundamental domain
of the orbifold to that of the torus, i.e.\ 
$-R_4/4 \leq x_4 \leq 3R_4/4$. The gauge 
functions can then be extended by the action of the orbifold group. By
combining the transition functions in the two copies of the orbifold
fundamental domains, one finds that  $x_4 \rightarrow x_4 + R_4$   is
accompanied by the gauge transformation $h^{-1}_0 \Omega h_0 \Omega$. If
$\Omega$ acts non-trivially on the $SU(2)$  (i.e.\ $\Omega$ is not in the
centre of $SU(2)$) then this is equivalent to turning on an $SU(2)$ Wilson
line along the $x_4$ direction. Such a Wilson line has the effect of 
introducing 
different holonomies around different fixed points \cite{inq}. In the present
example, this amounts to having holonomies $\Omega$ around the fixed points
$(0,0)$ and $(0,R_2/2)$ and $h^{-1}_0 \Omega h_0$ around the other two fixed
points. The fact that the gauge transition function $h_0^{-1}$ introduces
these different holonomies can also be seen by considering the eigenvalue
problem for the Laplacian in the $x_4,x_5$ directions.

Since we have realized the orbifold group action as a
$\Zint_2$ shift, the only non-trivial such $\Omega$ is necessarily of the form
$\Omega = i\sigma_3 \Omega'$, where $\Omega'$ acts on $E_7$ and corresponds to 
a shift by half the {\bf 56}-weight. Note that for this choice of $\Omega$
the roots of $SU(2)$ give rise to hypermultiplets and, as a result, 
there is no $SU(2)$ enhancement in the four-dimensional theory; instead, 
what appears at this point is massless hypermultiplets. 
If $h_0$ is not in the $U(1)$ generated by $\sigma_3$, 
then there is a non-trivial Wilson line that can be parametrized by a
suitable choice of basis as $e^{i\theta \sigma_1}$, where $\theta$ is a
continuous parameter. For $\theta=0$ mod $2\pi$ there is no Wilson line
and the resulting theory is the same as the original one before turning on
the flux. Non-zero values of $\theta$ correspond to higgsing the original
theory by giving non-zero expectation values to hypermultiplets 
corresponding to the $SU(2)$ roots. 

In fact, for this example, one can construct an explicit CFT description of
these models. One can choose a basis where the Wilson line is in the Cartan
direction (say $e^{i\theta \sigma_3}$). In this basis the orbifold action
$\Omega = i \sigma_1$ reflects the Cartan direction. In other words,
one obtains an asymmetric orbifold \cite{nsv}, where the $\Zint_2$ action
is defined by reflection of $T^4$ together with reflection of the chiral
boson $\phi$ corresponding to the Cartan direction of $SU(2)$, and a shift
in the $E_7$ given by half the {\bf 56}-weight.
Clearly this $\Zint_2$ is an automorphism of the lattice deformed by the
Wilson line: since the latter is along $x_4$ in the Cartan direction of
$SU(2)$, the deformation of the lattice involves the $x_4$ and $\phi$
directions and, as a result, reflecting $x_4$ and $\phi$ simultaneously
leaves the lattice unchanged. Moreover, it is easy to see that this model
satisfies the level matching condition: the original shift of half the
fundamental weight of $SU(2)$, as well as the reflection of $\phi$,
contribute 1/16 to the level matching condition. In this CFT description,
the parameter $\theta$ corresponds to higgsing by giving a vacuum expectation
value to the Wilson line
that, as a result of orbifold group projection, is part of a hypermultiplet.  
Thus the new supersymmetric vacuum is in the same class as that of the original
vacuum before turning on the magnetic flux. In fact this is not surprising:
in the original supersymmetric vacuum, the special points in the vector
moduli space, where the untwisted charged states become massless, the
latter come in pairs. As a result there is always a flat direction in the
Higgs branch. Now after turning on a field strength, if there is a new
supersymmetric minimum at non-zero expectation value of the charged
states, then clearly the new vacuum is in the original Higgs branch. 

Since in the four-dimensional theory the flat $T^2$ was a spectator in
this entire analysis (apart from providing the Coulomb phase), one can
decompactify this to a six-dimensional theory. In this case the original
orbifold model gives an $E_7\times SU(2)$ gauge group 
with charged hypermultiplets from the untwisted sector in the {\bf (56,2)}
representation. In the new supersymmetric vacuum, the spectrum can be analysed
for different values of the parameter $\theta$. For $\theta=0$ one has the
original spectrum, while for $\theta=\pi$ the gauge group is broken to
$E_6\times U(1)^2$ and for generic $\theta$ the gauge group is $E_6 \times
U(1)$, which has lower rank. One can also study the spectrum of massless
hypermultiplets and, as expected, the theory is anomaly-free for all values of
$\theta$. In fact the point $\theta = \pi$  could have been described in
the original basis as an orbifold in the presence of Wilson lines, as studied 
in Ref.\ \cite{inq}.
\vskip 1cm

\noindent {\bf $\Zint_4$ orbifold}

In the $\Zint_2$ example considered above, at the special point in the moduli 
space where extra massless charged hypermultiplets appeared, they  
did so in pairs, and as a result
provided a flat direction. The new vacuum then turned out to be in the
same class as the old vacuum before turning on the magnetic field. It is
instructive to consider another orbifold example where only one charged
massless hypermultiplet appears. In this case there is no flat direction 
that would break the $U(1)$ in question. In the presence of the magnetic field,
if there is a new supersymmetric vacuum, it should be either the same
as the original one or a different one, disconnected from the
original. This example would therefore be closer in spirit to the 
one considered in Ref.\ \cite{ps} on the type II side.

Such examples are provided by $\Zint_3$ or $\Zint_4$ orbifolds. For notational
simplicity we consider here a $\Zint_4$ orbifold, which is realized by 
simultaneous $\pi/2$ rotation in the planes 
$(x_4,x_5)$ and $(x_6,x_7)$, together with a shift given by one fourth the
weight of {\bf (2,56)} in the decomposition of 
$E_8$ in terms of $SU(2)\times E_7$. 
For this orbifold, the gauge group is broken to $U(1)\times E_7$, with several
hypermultiplets transforming under different representations of the gauge group.
The hypermultiplets that we are interested in are the ones that are
neutral under $E_7$ and charged under $U(1)$. In the untwisted sector, this
is provided by the $SU(2)$ root and, contrary to the $\Zint_2$ case, there is
just one such hypermultiplet, which we denote by $\Phi$. 
We can further go to the Coulomb phase by turning on Wilson lines 
along the $(x_8,x_9)$ directions so that $E_7$ is broken to
$U(1)^7$ and all the charged hypermultiplets, including $\Phi$, become massive.

Let us now turn on a magnetic field in the $(x_4,x_5)$ plane along the first 
$U(1)$. The quantization condition can be found, as before, by going to the 
fundamental domain defined by $0 \leq x_4 \leq R/2$  and $0 \leq x_5 \leq R/2$ 
(the two radii are equal as required by $\Zint_4$ symmetry), see Fig.\ 1. The
two fixed points in this domain are $(0,0)$ and $(R/2,R/2)$. We can
choose two patches as follows: patch I is the lower triangle with vertices
$(0,0)$, $(0,R/2)$ and $(R/2,0)$ and patch II is the complementary triangle. 
Each of these patches is topologically a disc, owing to the $\Zint_4$
identification of the boundaries: in patch I, the boundary $(x,0)$ for $0 \leq
x \leq R/2$ is identified with the boundary $(0,x)$; in
patch II, the boundary $(R/2,x)$ is identified with the one defined by
$(x,R/2)$. In other words, various fields on these patches are identified
on these boundaries up to a gauge transformation $\Omega$ corresponding 
to the $\Zint_4$ shift in the gauge sector defined above (we are
considering only fields that do not depend on $x_6$ and $x_7$, otherwise
$\Omega$ will be accompanied by the orbifold group action on 
these variables). The $U(1)$ gauge potential that gives rise to a
constant magnetic field can be defined on the two patches, such that it 
satisfies the boundary identifications, as
\begin{equation}
A_4^I = B x_5 \sigma_3 ~~~~~~~~~~A_5^I = -B x_4 \sigma_3
\label{z4aI}
\end{equation}
in patch I and
\begin{equation}
A_4^{II} = B \left(x_5-{R \over 2}\right) \sigma_3 ~~~~~~~~~~
A_5^{II} = -B \left(x_4-{R\over 2}\right) \sigma_3
\label{z4aII}
\end{equation}
in patch II. In the overlap between the two patches, the gauge transition
function is $g_{II,I} = \exp [iBR(x_4-x_5) \sigma_3/2]$. The identification
of the points $(0,R/2)$ with $(R/2,0)$, and the fact that 
the spectrum includes the fundamental representation of $SU(2)$, 
then gives the quantization condition $B= 4\pi n/R^2$ for some 
integer $n$.

At the special points in the vector moduli space where the
hypermultiplet $\Phi$ defined above corresponding to $SU(2)$ roots
becomes massless, we can perform a gauge transformation $g_I$ in, say, patch
I with $g_I = h g_{II,I}$ at the overlap of the two patches, and at
the boundaries $g_I(0,x) = g_I(x,0)
= h g_{II,I}(0,R/2)$, where $h$ is a constant $SU(2)$ group element.  
It is clear that such a smooth $g_I$ exists owing to the quantization
condition of $B$. The identification of the boundaries $g_I(0,x)=
\Omega g_I(x,0) \Omega^{-1}$ implies that $h$ commutes with $\Omega$
and as a result $h$ is in the $U(1)$ subgroup generated by $\sigma_3$.
The new transition function $g'_{II,I} = g_{II,I} g_I^{-1} =h$ is constant
and therefore admits a zero gauge potential in the two patches. Thus we
see that once again at the special point where the charged hypermultiplet 
$\Phi$ becomes massless (and hence tachyonic in the presence of a magnetic 
field) one can smoothly take the field strength to zero, thereby obtaining 
a new supersymmetric vacuum.

In order to understand the nature of the new vacuum, we must consider the
holonomies around the fixed points. Just as in the $\Zint_2$ case, 
either by extending the fundamental domain to the original torus or by
analysing the eigenvalue problem of the Laplacian on the fundamental
domain, we can see that if the holonomy around the fixed point $(0,0)$
is $\Omega$, then around the fixed point $(R/2,R/2)$ it is
$h^{-1} \Omega h$, which is equal to $\Omega$ since $h$ commutes with $\Omega$.
Therefore the new vacuum corresponds to having no
Wilson lines in the $(x_4,x_5)$ plane. Thus we conclude that the new
vacuum is in fact the original supersymmetric vacuum before turning on
the magnetic field.

Although all of the above discussion was carried out for the $N=2$ theory, 
it can be
extended to $N=1$ orbifold models. For example, one can consider a $\Zint_2 
\times \Zint_2$ orbifold, with the first $\Zint_2$ as defined above and
the second  $g$ acting on $x_6$, $x_7$, $x_8$ and $x_9$ together with a
$\Zint_2$ shift along $x_4$, $x_5$ directions satisfying the level-matching 
condition. The tachyon now appears in the $N=1$ chiral multiplet,
which is accompanied by higher Landau levels.
It is clear that the analysis given for the $N=2$ case remains unchanged and 
the new
supersymmetric vacuum can again be given a CFT description as a $\Zint_2
\times \Zint_2$ orbifold. This shows that this mechanism of supersymmetry 
restoration, at least in some cases, also works for $N=1$ theories. It
will be interesting to study this phenomenon in more general situations 
as well as in the type II duals.

\section {\bf Tachyons without Landau Levels}

Let us now consider the tachyons coming from the twisted sector in heterotic
$N=2$ orbifolds, which, as we have noticed in Section 2, do not give rise
to towers of Landau levels with masses of order $g_H^2/V$. Therefore the
analysis of the potential to the leading order in $1/V$ can be carried out by
restricting to the quartic terms in the tachyons. Moreover the quartic terms
that enter in this analysis are zeroth order in $1/V$ and therefore can be 
calculated in the supersymmetric theory by setting the magnetic flux to zero. 
Assuming that the tachyons are charged only under the $U(1)$ gauge generator 
along which we have the magnetic flux, these quartic terms
are given by the $N=2$ D-terms:
\begin{equation}
{1\over 2} g_H^2\, \left\{\bigg[ \sum_a q_a (|\chi_a|^2-|\eta_a|^2)\bigg]^2
+4\bigg|\sum_a q_a\chi_a{\bar\eta}_a\bigg|^2 \right\},
\label{Dterm}
\end{equation}
where the index $a$ runs over the different tachyons and $q_a$ are the absolute
values of their charges. 

In Section 2, we discussed the mass terms for tachyons 
in type I theory arising from
95 strings. In the type I string frame they are given by Eq.\ (\ref{levelI}).
In order to obtain the corresponding mass formula for the heterotic theory, we
have to use the duality relations \cite{abfpt}:
\begin{equation}
V_I={V_H\over g_H} \omega_H^{-1/2}\qquad ;\qquad
g_I^2=g_H \omega_H^{-1/2}\, ,
\label{durel}
\end{equation}
where $V_I$ and $V_H$ are the volumes of the tori on which the magnetic flux is
turned on, in type I and heterotic theories, respectively; $\omega_H$ is the
total volume of the six-dimensional internal space in the heterotic
theory; $g_I$ and $g_H$ are the four-dimensional respective string
coupling constants. Starting from the mass formula (\ref{levelI}) and
going to the Einstein frame, while using the quantization condition for the
magnetic field $F=2\pi/V_I$ in units where we normalize the minimum charge to
1, we obtain the tachyon mass squared $M^2=-|q|g_I^2/V_I$. 
This is mapped in the heterotic theory,
by means of the duality relations (\ref{durel}), to $M^2=- |q| g_H^2/V_H$.
Combining this mass term with the quartic potential (\ref{Dterm}) and the
cosmological constant obtained previously in Eq.\ (\ref{pot}) 
by integrating $F^2$ over the internal space, we obtain:
\begin{equation}
{\cal V}_{\rm eff}={1\over g_H^2}\, \bigg({2\pi\over V_H}-\sum_a q_a|\chi_a|^2
\bigg)^2\ ,
\label{Vtwist}
\end{equation}
where we have set to zero all the non-tachyonic components $\eta_a$'s.

The above potential has a minimum at $\sum_a q_a|\chi_a|^2=(2\pi/ V_H)^2$, at
which the cosmological constant vanishes and supersymmetry is restored.
Unlike in the previous case, the $U(1)$ is spontaneously broken at the new 
minimum, with its gauge field absorbing one linear combination of
hypermultiplets. This is analogous to the type II situation analysed in 
Ref.\ \cite{ps}.

\section {\bf Conclusions}

We would now like to see what the results on the heterotic side imply for
the type IIA side. In particular, the first case involving Landau levels
studied here should have a counterpart on the type IIA side. In fact, it
is not hard to see what these states are since they have a six-dimensional
interpretation. They are the 2-branes wrapped around 2-cycles in the $K_3$
fibre (which survive the fibration) and carry Kaluza--Klein momenta along the
base. Thus these are the original 2-branes boosted along the directions of the
base. In the $N=4$ context, these are again BPS states that carry, besides the
charge associated to the 2-cycle in $K_3$, also the charges corresponding to
the Kaluza--Klein $U(1)$'s coming from the base $T^2$. In fact, these BPS states
are needed for the heterotic--type IIA duality in order to fill the $O(6,22)$
lattice, and their masses squared due to the Kaluza--Klein momenta 
(in the  Einstein frame) become of order  $A+g_{II}^2/V_{II}$, where $A$ is the
area of the 2-cycle. The limit that we are considering corresponds to 
$A \rightarrow 0$. 
In the $N=2$ case,
however, since there are no such Kaluza--Klein $U(1)$'s as the base is $S^2$,
they are non-BPS states. The masses squared of these 
states, as in the $N=4$ case, 
are shifted by an amount of the order of $g_{II}^2/V_{II}$. In the context of
Calabi--Yau spaces, these non-BPS 2-branes should appear through non-holomorphic
2-cycles \cite{bbs}. Of course, these states are in general unstable; 
however, they do contribute to the higher-dimensional terms in the 
effective potential for the BPS states. 

This is exactly what happened in the discussion in Section 3 for the 
heterotic case. There the quartic coupling was of
order $g_H^2$, while the masses squared were of order $g_H^2/V_H$. As a result, 
for example, one gets a reducible contribution at the 6-point level 
(with one KK
exchange) that behaves as $g_H^2 V_H$. On the type IIA side the quartic
coupling is $1/V_{II}$, and a similar reducible contribution at the 
$(2n+4)$-point
level due to these non-BPS states is (in the $A\rightarrow 0$ limit) 
of order $1/(V_{II}g_{II}^{2n})$. This
peculiar problem appears because we are taking the limit where $A$ is much
smaller than the coupling constant $g_{II}^2$ and, as a result, there is a
tower of very light non-BPS states. Thus, in the problem of minimization of
the potential in the presence of a magnetic field, these states must also be 
taken
into account. As in the heterotic case, this problem is more easily analysed 
by going to $D=6$. 

For example, let us take the (11,11)-model \cite{fhsv}, which is obtained 
as a $\Zint_2$ orbifold of $K_3 \times T^2$, with the magnetic
field turned on the $T^2$. In this model
there are two types of special points: one class with an $SU(2)$
enhancement along with an adjoint hypermultiplet; and a second class with an
$SU(2)$ enhancement along with four fundamental hypermultiplets. 
On the heterotic side the
first case appears in the untwisted sector, while in the second case the
fundamental hypermultiplets arise in the twisted sector. 
In particular, this means that 
the charged massless states that appear in the first case have a six-dimensional
interpretation. In fact on the type IIA side these states correspond to 
2-brane wrapped around one of the vanishing 2-cycles of $K_3$, which is even 
under the $\Zint_2$ Enriques involution. From the six-dimensional point of view
they appear as point particles. Upon further compactification to
$D=4$, these particles will in general carry Kaluza--Klein momenta
and provide higher Landau levels in the presence of a magnetic field. From
the ten-dimensional point of view, they are 2-branes wrapped around vanishing
2-cycles of $K_3$ that are boosted along the base. In this case, we can
analyse this problem exactly as in
Section 3, by going to the six-dimensional theory on $R^4 \times T^2/\Zint_2$, 
and we find that the new supersymmetric minimum corresponds
to the old vacuum up to non-trivial field redefinitions.    
In the second case, if we use the tachyons in the fundamental representation,
which do not have higher Landau levels in the heterotic theory, 
then the analysis of Section 4 shows that, in the new supersymmetric vacuum, 
$SU(2)$ is completely broken by absorbing three hypermultiplets. Thus, in the
new vacuum relative to the generic points in the old one, the number of vectors
decreases by 1 while the number of hypermultiplets increases by 5.

In the $N=2$ case, there are also charged states that have no six-dimensional 
interpretation in the large volume limit of the base. It is simpler to study
these states on the heterotic side compactified on $K_3\times T^2$.
In the orbifold limit, these states come from the twisted sector and do not
produce tachyons with Landau levels since they do not carry $K_3$ momentum.
However, on a smooth $K_3$ manifold obtained by switching on the blowing-up
modes $\cal B$ one expects them to have $K_3$ Kaluza--Klein excitations in 
the limit of 
volume of the base $V_H\gg {\cal B}^{-2}$. By duality the same behaviour 
would therefore be expected for type II on Calabi--Yau, which is a $K_3$ 
fibration. 
Hence, apart from special points such as orbifolds, we expect that in a generic
compactification there are Landau levels. A correct treatment should therefore
also include these states.
\vskip 1cm

\noindent {\bf Acknowledgements}

We would like to thank M. H. Sarmadi and G. Thompson for useful
discussions. T.R.T.\ acknowledges the hospitality of the Ecole Polytechnique at
the initial stage of this work and of the CERN Theory Division
during its completion.

\end{document}